\documentclass[a4paper,twocolumn]{article}
\pdfoutput=1

\usepackage[pdftex]{graphicx}
\usepackage{latexsym}
\usepackage{url}
\usepackage{comment}
\usepackage{amsmath}
\usepackage{pifont}
\usepackage{times}
\usepackage{newtxmath}
\usepackage{authblk}
\usepackage{xspace}

\usepackage{CJKutf8}

\newcommand{\cmark}{\ding{51}}%

\newcommand{\etal}{\textit{et al.}}%

\begin{document}
\title{A Survey of Refactoring Detection Techniques\\Based on Change History Analysis%
  \footnote{%
    This article is a private translation of the article published in the JSSST journal \textit{Computer Software}:
    Choi, E., Fujiwara, K., Yoshida, N., and Hayashi, S.: A Survey of Refactoring Detection Techniques Based on Change History Analysis, Vol.~32, No.~1(2015), pp. 47--59.
    The electronic copy of the original version can be obtained from\protect\\\protect\url{http://doi.org/10.11309/jssst.32.1_47}.
  }%
  \footnote{%
    Notice for the use of this material: The copyright of this material is retained by the Japan Society for Software Science and Technology (JSSST)\@.
    This material is published on this web site with the agreement of the JSSST\@.
    Please be complied with Copyright Law of Japan if any users wish to reproduce, make derivative work, distribute or make available to the public any part or whole thereof.
  }%
}
%
\author[1]{Eunjong Choi\footnote{The author is currently with Graduate School of Science and Technology, Nara Institute of Science and Technology, Japan. Email: choi@is.naist.jp}}
\author[2]{Kenji Fujiwara\footnote{The author is currently with National Institute of Technology, Toyota College, Japan. Email: fujiwara@toyota-ct.ac.jp}}
\author[3]{Norihiro Yoshida\footnote{The author is currently with Center for Embedded Computing Systems, Graduate School of Informatics, Nagoya University, Japan.  Email: yoshida@ertl.jp}}
\author[4]{Shinpei Hayashi\footnote{The author is currently with School of Computing, Tokyo Institute of Technology, Japan. Email: hayashi@c.titech.ac.jp}}
\affil[1]{Graduate School of Information Science and Technology, Osaka University}
\affil[2]{Graduate School of Information Science, Nara Institute of Science and Technology}
\affil[3]{Graduate School of Information Science, Nagoya University}
\affil[4]{Graduate School of Information Science and Engineering, Tokyo Institute of Technology}
\renewcommand\Authands{ and }
\date{\vspace{-5ex}}%


\maketitle
\begin{abstract}
Refactoring is the process of changing a software system in such a way that it does not alter the external behavior of the code yet improves its internal structure.
Not only researchers, but also practitioners, need to know about past refactoring instances performed in a software development project.
So far, a number of techniques have been proposed for automatic detection of refactoring instances.
Those techniques have been presented in various international conferences and journals, however, it is difficult for researchers and practitioners to grasp the current status of studies on refactoring detection techniques.
In this survey paper, we review various refactoring detection techniques, especially techniques based on change history analysis.
First, we give the definition and categorization of refactoring detection methods in this paper, and then introduce refactoring detection techniques based on change history analysis.
Finally, we discuss possible future research directions for refactoring detection.
\end{abstract}

\section{Introduction}

Refactoring is the process of changing a software system in such a way that it does not alter the external behavior of the code yet improves its internal structure~\cite{Fowler1999,opdyke1992}.
Refactoring is performed for various reasons~\cite{Fowler1999}.
For example, it can help prevent the introduction of new defects into source code by improving the maintainability of source code with high complexity or low readability.
Not only researchers, but also practitioners, are interested in detecting refactoring instances, and many books and papers on refactoring detection techniques have been published~\cite{Fowler1999,kerievsky2005,Mens2004}.

Both practitioners and researchers need to know about past refactoring instances performed in a software development project.
Their needs:

%
\begin{itemize}
\item Practitioners want to use refactoring information to determine whether and how to track software to be maintained, by understanding the refactoring implemented on libraries, frameworks and Application Programming Interfaces (API) being used ~\cite{dig-ecoop2006,taneja-ase2007}.
\item Researchers want to collect this information to conduct empirical studies of refactoring and its effects and to support techniques by collecting refactoring instances~\cite{Choi2014,Kim2011}.
\end{itemize}

However, when refactoring changes are saved together with non-refactoring changes, it takes much time to determine whether the source code was modified by refactoring~\cite{ge-chase2014,hayashi-wcre2013,murphy-hill-tse-2012}.
Since large-scale software projects often have thousands of modifications in their change histories, it is difficult to check whether refactoring was performed by manually analyzing all the changes.

Several techniques for automatically detecting refactoring instances, hereafter referred to as `refactoring detection techniques', have been proposed.
These techniques are published in various journals and international conferences, making it difficult to review all of these techniques.
In 2004, several papers surveying the research and techniques for refactoring detection were published~\cite{Bois2004,Mens2004,Mens2003}. 
However, since 2004, many additional papers on refactoring detection techniques have been published. Therefore, it is difficult to grasp the current trend of research on refactoring detection techniques from the 2004 survey papers.

In this paper, we introduce refactoring detection techniques based on change history analysis.
Section~\ref{sec:definition} defines the refactoring terms used in this paper. 
Section~\ref{sec:category} classifies refactoring detection techniques, and then introduces the techniques based on change history analysis.
Section~\ref{sec:main} presents refactoring detection techniques based on change history analysis of artifacts, while Section~\ref{sec:discussion} discusses directions for future research.
Section~\ref{sec:summary}, finally, concludes the paper with a brief summary.

\section{Definition of Refactoring Detection Terms}
\label{sec:definition}

In general, details of refactoring instances are listed in refactoring catalogs.
Each entry in a refactoring catalog includes the preconditions, postconditions, detailed procedures, and other parts of a refactoring operation, along with the name of the refactoring operation.
Some catalogs describe refactoring operation as a software pattern, in which case these patterns are called refactoring patterns.
For example, the refactoring operation that moves a method belonging to a class into another class is called the \textit{Move Method}, and this pattern is listed in a catalog along with the preconditions to perform it and other parts of the operations.
Refactoring catalogs are usually published as books or on the web~\cite{FowlerWeb,Fowler1999,kerievsky2005}.

In this paper, we use the refactoring detection terms.
We also define refactoring detection as follows.
When a pair of versions extracted from the version sequence of a software product $(v_{0}, \dots, v_{n-1}, v_{n})$ is given as $(v_{a}, v_{b})~(0 \leq a<b \leq n)$,
we denote the changes from $v_{a}$ to $v_{b} $ as $C =\{c_{0}, \dots, c_{m-1}, c_{m}\}$.
In this case, we define refactoring detection as inferring whether a refactoring operation, included in a refactoring catalog, is contained in the non-empty subset of the change set $C$.
In general, tools for refactoring detection output information such as ``\textit{Pull up Method} and \textit{Move Field} are performed from version $v_{1}$ to $v_{2}$'' when the pair of versions $(v_{1}, v_{2})$ are input.
However, there are some tools that do not output specific refactoring names, but just suggest the existence of refactoring instances~\cite{soares-sbes-2011}.

A refactoring instance does not always exist as an individual change in a pair of versions, but exists along with other modifications~\cite{hayashi-ieicet2010,murphy-hill-tse-2012,Parnin:2006}.
G\"{o}rg and Wei{\ss}gerber called a change that contains mixed with refactoring and non-refactoring modifications \textit{impure refactoring}~\cite{gorg-iwpc2005}.
Unlike \textit{pure refactoring}, a pair of versions related by \textit{impure refactoring} does not always keep the external behavior fixed.
Murphy-Hill \etal\ pointed out that refactoring is often performed while also adding features and fixing bugs, and they call such refactoring \textit{floss refactoring}~\cite{murphy-hill-tse-2012,Murphy-Hill2007}.
Compared with \textit{root-canal refactoring} that distinguishes refactorings from other changes, \textit{floss refactoring} often generates pairs of versions that include both refactoring and other non-refactoring modifications.
Murphy-Hill \etal\ reported that \textit{floss refactoring} is often performed~\cite{murphy-hill-tse-2012}.
Also, Herzig and Zeller reported that there are \textit{tangled code changes} containing various kind of changes~\cite{herzig-msr2013}.

Since compound changes often occur in real software development as mentioned above, it is necessary for refactoring detection techniques to detect the performance of refactoring despite the changes from version $v_{a}$ to $v_{b}$ including not only refactoring operations, but also bug fixes and/or feature additions~\cite{hayashi-ieicet2010,murphy-hill-tse-2012}.
In this paper, we include refactoring detection for these compound changes in our survey.

\newcommand{\BL}[1]{\begin{minipage}{10em}\centering#1\end{minipage}}
\begin{table*}[t]
\centering
\caption{Research methods of refactoring detection techniques}
\label{table:detection_approaches}
\begin{tabular}{c|ll}
	 & \multicolumn{1}{c}{context} & \multicolumn{1}{c}{fidelity} \\ \hline
explicit & \textbf{A1}: commit log mining & \textbf{A3}: tool usage logs \\
implicit & \textbf{A2}: developer observation & \textbf{A4}: analyzing histories
\end{tabular}
\end{table*}

Refactoring detection techniques share their technical background with several differential analysis techniques.
For example, research on adding a well-known name to a set of changes, such as \textit{systematic change} detection has been conducted~\cite{Kim2009,Kim2007}.
Moreover, several origin analyses that identify the correspondence between code fragment in a certain version and in previous version include techniques that recognize when the name of a program entity is replaced~\cite{Godfrey2005,Kim2005}.
Similarly, some techniques used to recognize comprehensive differences in source code or software models have analysis methods similar to those used in refactoring detection~\cite{Fluri2007}.
In this paper, we do not cover all these differential analysis techniques because our main purpose is investigating techniques that detect refactoring operations listed in refactoring catalogs.
As an exception, we do survey techniques used to verify consistency of program behavior for refactoring detection, even though these techniques do not identify concrete refactoring patterns.

\section{Categorization of Refactoring Detection Techniques}
\label{sec:category}

This section classifies the refactoring detection techniques described in the previous section into four different research methods, and then describes the target of this survey, refactoring detection techniques based on change history analysis.

Murphy-Hill \etal\ categorized refactoring detection techniques into four research methods based on two perspectives, context versus fidelity and explicit versus implicit information~\cite{murphy-hill-wrt-2008}.
Table~\ref{table:detection_approaches} shows their four research methods.
For one of their axes, their categorization depends on whether or not refactoring instances are identified by using explicit information about refactoring events (\textit{explict} or \textit{implicit}).
For the other axis, their categorization uses whether refactoring instances are determined by subjective judgments or observable facts (\textit{context} or \textit{fidelity}).
Next, we discuss the details of each of these four research methods.

First, the \textbf{A1}: commit log mining set of techniques identify refactoring by analyzing the commit logs of version control systems~\cite{oba-2013,ratzinger-msr-2008,stroggylos-wosq-2007}.
If a developer has noted the performance of refactoring in the commit log, refactoring instances are identified by extracting the correct log entry.
Therefore, these techniques search for words expressing refactoring activities such as `refactor' or `extract' in the commit logs.
A characteristic of these techniques is the use of explicit records of refactoring performed by developers. However, the accuracy of identifying refactoring performance and its descriptions depends highly on the subjective judgment of the developer.
These techniques can be applied to any software system which has a history of software development using a version control system.
However, a disadvantage of this method is that the accuracy of the refactoring information depends on the judgment of the developers, and the location of performed refactoring may be missed.
Murphy-Hill \etal\ compared commit logs of version control systems with performed refactoring and found that commit logs contain unreliable information of refactoring~\cite{murphy-hill-tse-2012}.
Therefore, when researchers use this method to investigate the refactoring performed by developers, they should take into consideration that this method provides biased information about refactoring.

Next, in the \textbf{A2}: developer observation set of techniques, researchers identify past refactoring by directly observing developers' works or by using screen capturing tools for indirect observation~\cite{boshernitsan-sigchi-2007,murphy-hill-icse-2008,pizka-serp-2004}.
A concrete example of this set of techniques is a technique that periodically captures developers' screen activities while doing software development using a tool, and then identifies refactoring instances from the recorded information.
The records used in this method do not provide explicit information about the performed refactorings.
Moreover, since the researcher determines whether a developer conducted refactoring or not, the refactoring information in this method is based on the subjective judgments of the researcher.
Although the applicability of this method is limited, it provides detailed information about development histories.

Third, the techniques classified as \textbf{A3}: tool usage logs identify refactoring operations by collecting the logs of refactoring support tools~\cite{dig-icse-2007,murphy-hill-ieee-softw-2006,robbes-icse-2008}.
These refactoring support tools enable the automatic application of representative refactoring patterns in an integrated development environment.
It is obvious that developers adopt these tools in order to conduct refactoring.
This method can collect information of \textit{pure refactoring} since refactoring support tools guarantee the preservation of external behavior. However, this method only captures certain kinds of refactoring patterns that are supported by the refactoring support tools.

Finally, techniques classified as \textbf{A4}: analyzing histories identify refactoring instances by analyzing a sequence of versions of software development artifacts.
In this method, refactoring is not determined by the subjective judgments of developers or researchers, but by observable facts based on changes in the artifacts such as source code.
However, this method might miss past refactoring instances performed by developers.

In this paper, we mainly introduce techniques in the \textbf{A4} method.
Recently, recording software change histories has become very popular in software development companies and open source software projects.
Therefore, these techniques can be widely applied, and it is expected that more refactoring instances can be identified than with the techniques classified as \textbf{A1} to \textbf{A3}.
This paper focuses on these techniques as a target because they can also identify refactoring instances applied to libraries and frameworks, and these techniques support empirical research on refactoring and its impacts.

\section{Refactoring Detection Techniques}
\label{sec:main}
\subsection{Techniques Based on Change History Analysis}

In this study, we selected techniques based on change history analysis as the target from the four methods of techniques described in Section~\ref{sec:category}.
This set of techniques was selected because change history analysis can identify more refactoring instances, as compared to the other techniques, because performed refactoring always remains in the history.

Furthermore, we investigated papers on refactoring detection techniques based on change history analysis that have been published in major international conferences on software engineering (APSEC, ASE, CSMR, FSE, ICSE, ICSM, MSR, OOPSLA, SCAM, and WCRE) and journals (IEEE Transactions on Software Engineering, Information and Software Technology, Journal of Systems and Software, and Journal of Software: Evolution and Process) and then analyzed their approaches.
Based on our results, we categorized the techniques into the following six types:

\begin{itemize}
  \item Rule-based approach,
  \item Code clone analysis-based approach,
  \item Metrics-based approach,
  \item Dynamic analysis-based approach,
  \item Graph matching-based approach, and
  \item Search-based approach.
\end{itemize}

Table~\ref{tab:category} categorizes a number of refactoring detection approaches based on change history analysis into these six approaches.
The approaches introduced in this paper were selected from papers published in the target international conferences and journals based on the importance of the papers in these categories.
We included all the important papers, as far as we know.
Next, in Sections~\ref{sec:rule} to \ref{s:search-based}, we describe the details of these on refactoring detection techniques in these categories.
Note that techniques that use multiple approaches to detect refactoring instances may be included in multiple categories.

\newcommand{\R}[1]{\rotatebox{90}{#1}}
\begin{table*}[t]\centering\small
\caption{Categorization of refactoring detection approaches}
  \begin{tabular}{c|l||c|c|c|c|c|c}
    Year &                                                 & Rule        & Code clone analysis & Metrics     & Dynamic analysis     & Graph matching     & Search      \\ \hline \hline
    2000   & Demeyer~\cite{demeyer-oopsla2000}                    &              &                 & \cmark         &              &                      &           \\ \hline
    2004   & Antoniol~\cite{antoniol-iwpse2004}                   & \cmark       &                 &                &              &                      &           \\ \hline
    2005   & G\"{o}rg~\cite{gorg-iwpc2005}                        & \cmark       &                 &                &              &                      &           \\ \cline{2-8}
           & Xing~\cite{xing-ase2005,xing-wcre2006,xing-ase2007}  & \cmark       &                 &                &              &                      &           \\ \hline
    2006   & Advani~\cite{advani-sac2006}                         & \cmark       &                 &                &              &                      &           \\ \cline{2-8}
           & Wei{\ss}gerber~\cite{weissgerber-ase2006}            & \cmark       & \cmark          &                &              &                      &           \\ \cline{2-8}
           & Dig~\cite{dig-ecoop2006}                             & \cmark       &                 &                &              &                      &           \\ \hline
    2007   & P\'{e}rez~\cite{perez-evol2007}                      &              &                 &                &              & \cmark               & \cmark    \\ \cline{2-8}
           & Taneja~\cite{taneja-ase2007}                         & \cmark       &                 &                &              &                      &           \\ \hline
    2008   & Hayashi~\cite{hayashi-apsec2008,hayashi-ieicet2010}  & \cmark       &                 &                &              &                      & \cmark    \\ \hline
    2010   & Kim~\cite{kim-fse2010}, Prete~\cite{prete-icsm2010}   & \cmark       &                 &                &              &                      &           \\ \hline
    2011   & Biegel~\cite{Biegel_msr2011}                         & \cmark       & \cmark          &                &              &                      &           \\ \cline{2-8}
           & Soares~\cite{soares-sbes-2011}                       &              &                 &                & \cmark       &                      &           \\ \cline{2-8}
           & Kehrer~\cite{kehrer-ase2011}                         &              &                 &                &              & \cmark               & \cmark    \\ \cline{2-8}
           & Thangthumachit~\cite{zui-apsec2011}                  &              &                 &                &              &                      & \cmark    \\ \hline
    2012   & Fadhel~\cite{fadhel-icsm2012}                        &              &                 &                &              &                      & \cmark    \\ \hline
    2013   & Mahouachi~\cite{mahouachi-ssbse2013}                 &              &                 & \cmark         &              &                      & \cmark    \\ \cline{2-8}
           & Soetens~\cite{soetens-icsm2013}                      &              &                 &                &              & \cmark               &           \\ \cline{2-8}
    	   & Fujiwara~\cite{fujiwara-fose-2013}                       & \cmark       &                 &                &              &                      &           \\
  \end{tabular}
\label{tab:category}
\end{table*}

\subsection{Rule-Based Approach}
\label{sec:rule}

\begin{figure}[t]
	\centering
	\vspace{1mm}
	\includegraphics[scale=0.48]{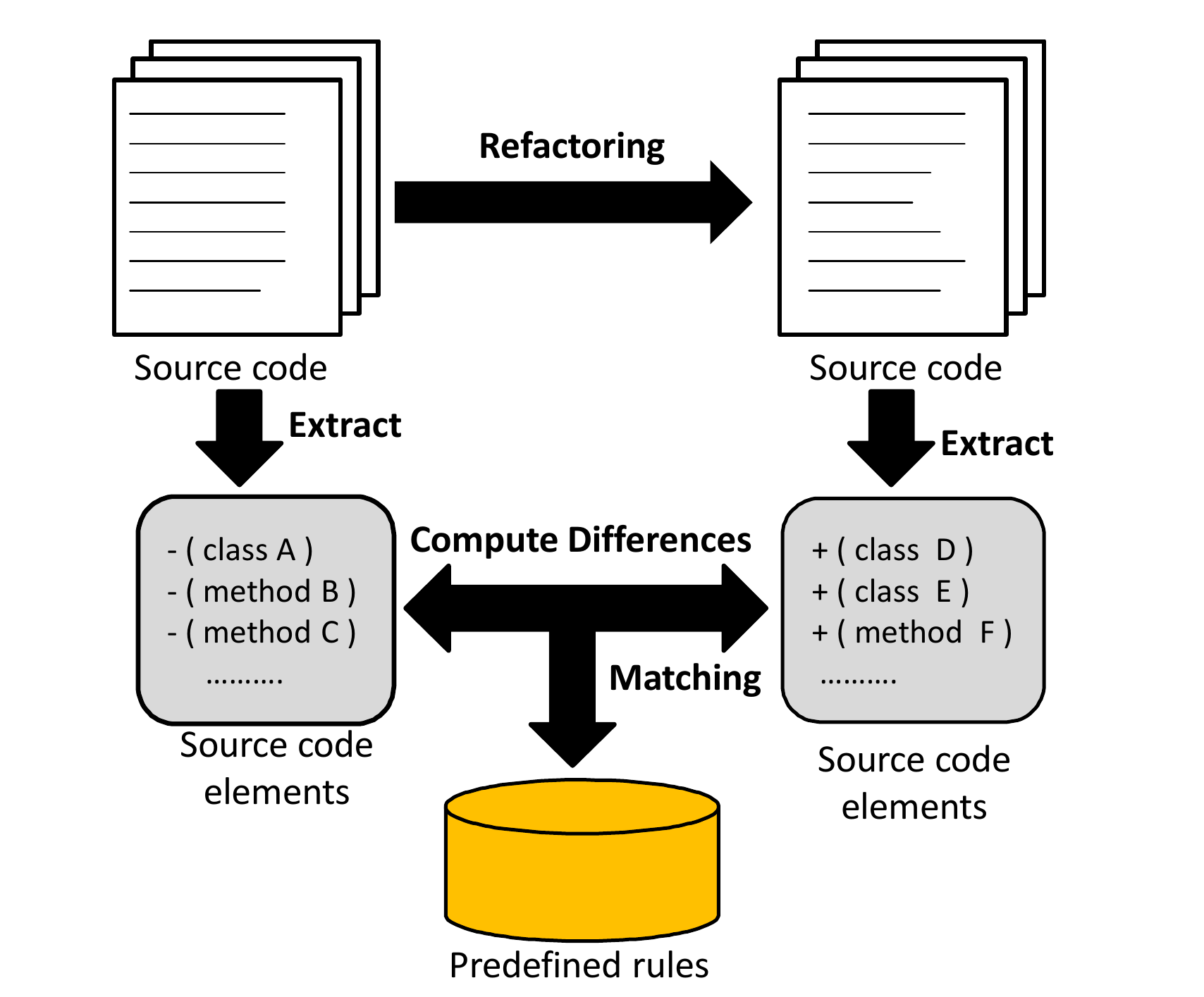}
	\caption{Process of a rule-based refactoring detection approach}
	\label{fig:rulebased}
\end{figure}

In this approach, rules are the criteria used to determine whether refactoring was performed based on changes, e.g., additions, deletions, and movements of the program elements, e.g., classes, methods, and parameters, and the similarity of the elements between two versions.
For example, to detect \textit{Extract Method} refactoring instances, a technique proposed by Prete \etal\ extracts program elements as facts and then computes the similarities in the facts between two versions~\cite{kim-fse2010,prete-icsm2010}. Then, if the computed results match a predefined rule that states the ``source code of a new method is extracted from a changed method in the old version, the new method calls the old method, and the source code of the new and old methods are similar to each other,'' then the target source code is detected as an \textit{Extract Method} refactoring instance.
Figure~\ref{fig:rulebased} summarizes the process of this detection approach.
The advantage of this approach is that it is easy to describe the detection rule for each refactoring pattern, because this approach can directly and declaratively express the changes between two versions.
However, the disadvantages of this approach are that the detection accuracy is low if the pre-defined rule is inadequate, and this approach is not suitable for detecting \textit{impure refactoring} instances containing both refactoring and non-refactoring changes.

The techniques proposed by Antoniol \etal\ and Advani \etal\ detect refactoring instances based on Fowler's definition of refactoring patterns~\cite{advani-sac2006,antoniol-iwpse2004}.
Antoniol \etal\ presented a technique for detecting refactoring instances at the class level, such as \textit{Class Extraction} and \textit{Class Split}, based on predefined conditions, which they used to investigate the evolution of classes in Java software systems.
Their technique extracts identifiers from each class, and then weights the extracted identifiers based on the Term Frequency-Inverse Document Frequency (TF-IDF).
Next, it converts the classes in each version into a vector based on the weights of the classes, and finally determines the refactoring instances according to the conditions based on the changes, e.g., a newly added class, in each class, along with the cosine of the angle between the two vectors representing the classes.
They applied this technique to 40 releases of \textit{dnsjava} and identified the \textit{Class Replacement}, \textit{Merge and Split}, and \textit{Factor Out} refactoring in these releases.

Advani \etal\ developed a tool for detecting refactoring instances according to predefined criteria aimed at investigating whether certain refactoring patterns are related~\cite{advani-sac2006}.
This tool reports refactoring instances when predefined criteria for 15 refactoring patterns are matched by changes in the class entities, e.g., methods and fields.
By applying this tool to seven open source software projects, this study found that the \textit{Rename Method}, \textit{Rename Field}, \textit{Move Method}, and \textit{Move Field} refactoring patterns are frequently related with other refactoring patterns.

G\"{o}rg and Wei{\ss}gerber implemented a tool called \textsf{REFVIS} for detecting refactoring instances based on changes, e.g., add, remove and unchanged, in the signatures of the classes and methods between two versions~\cite{gorg-iwpc2005}.
\textsf{REFVIS} also provides a feature that visualizes the detection results at the classes and methods levels.
Wei{\ss}gerber and Diehl presented a technique for detecting refactoring instances based on added, changed, or removed classes, interfaces, methods, and fields between two versions~\cite{weissgerber-ase2006}.
Their technique then ranks the refactoring instances based on similarities in the source code between the two versions using \textsf{CCFinder}, a token-based code clone detection tool.
This technique is able to detect similar sets of source codes as refactoring instances, whereas \textsf{REFVIS} only reports two exact matching sets of source code as refactoring instances.

Xing and Stroulia's \textsf{UMLDiff} detects refactoring instances between two versions~\cite{xing-ase2005,xing-wcre2006,xing-ase2007}.
\textsf{UMLDiff} extracts logical elements such as the types, names, and modifiers of the packages and classes from two input program versions.
It then computes their similarities based on changes, additions, movements, and deletions.
Finally, if the computed similarities are matched with rules representing a specific refactoring pattern, it identifies the target source code as a refactoring instance.

Prete \etal\ developed an \textsf{Eclipse} plug-in called \textsf{Ref-Finder} that detects refactoring instances of 63 refactoring patterns between two program versions based on predefined rules~\cite{kim-fse2010,prete-icsm2010}.
\textsf{Ref-Finder} extracts code elements, e.g., packages, classes, and interfaces, structural dependencies, e.g., containment and overriding relationships, and the contents of the code elements, e.g., if-then-else control structures, as elements from two input program versions.
It then computes the differences in the elements between the two program versions.
Finally, it determines the refactoring instances based on the predefined rules of the refactoring patterns~\cite{prete-tr2010}.
\textsf{Ref-Finder} detects both \textit{atomic refactoring} and \textit{complex refactoring} using other \textit{atomic refactoring} as a pre-requisite.
Furthermore, it can detect more refactoring patterns than \textsf{UMLDiff} by using code information such as conditional branch and exception handling.

A technique proposed by Fujiwara \etal\ detects refactoring instances in multiple revisions~\cite{fujiwara-fose-2013}.
This technique speeds up the refactoring detection by extracting code elements from each revision and matching them using Historage~\cite{hata-iwpse-evol-2011}.

A rule-based approach can also be used to detect refactoring instances from changes in histories of the components.
However, in general, because of backward compatibility where obsolete source codes coexist with their newer counterparts until the older code is, it is difficult to detect refactoring instances from changes in the histories of the components.
To address this problem, Dig \etal\ and Taneja \etal\ presented techniques for detecting refactoring instances between two versions of components based on predefined rules~\cite{dig-ecoop2006,taneja-ase2007}.
Dig \etal\ developed an \textsf{Eclipse} plug-in called \textsf{RefactoringCrawler}~\cite{dig-ecoop2006}, which identifies similar pairs of entities, e.g., methods and classes, in two versions of components.
This plug-in uses \textsf{Shingles}~\cite{broder-sequences1997} to find refactoring candidates, and then analyzes references among the source code entities in each of the two versions of the components to detect real refactoring instances.
Taneja \etal\ developed a tool called \textsf{Refac Lib}, which extracts similar entities from the source code from two API versions and then reports refactoring instances based on syntactic analysis, comparison of the similarities and sizes of the entities, and information regarding obsolete entities.

\subsection{Code Clone Analysis-Based Approach}
Research has also been done on the detection of refactoring using code clone detection tools that identify pairs or sets of duplicated code fragments in source code~\cite{Biegel:2010,Kamiya-2002}.
For refactoring detection, code clone detection tools can be used to identify moving and extraction of code fragments between versions.
The code clone analysis-based approach is able to identify code fragments that are not only identical code fragments but also code fragments that are slightly changed between versions.
It is difficult for this approach to detect refactoring instances that include various sorts of changes (e.g., \textit{impure refactoring}).

As mentioned in Section \ref{sec:rule}, Wei{\ss}gerber \etal~\cite{weissgerber-ase2006} proposed a technique to detect refactoring instances using a clone detection tool named \textsf{CCFinder}~\cite{Kamiya-2002}.
Their technique categorizes code changes between versions based on code similarity into three exclusive categories, EQUAL (exact match), CLONE (CCFinder-based match) and NONCLONE (all others).
The authors also investigated the relationship between these categories and refactoring.
Their results showed that the detection of EQUAL and CLONE cases increased precision and did not decrease recall.
Biegel \etal\ defined three types of code similarity, and then compared the performance of refactoring detection using these three types.
Two of their types were defined by either the similarity of token sequences and abstract syntax trees, or measured by the code clone detection tools \textsf{CCFinder} and \textsf{JCCD}~\cite{Biegel:2010}.
The other type was defined by string similarity, using \textsf{shingles}~\cite{broder-sequences1997} to represent the distance between strings.
Their investigation results showed that they could not confirm any significant difference between the three types of similarity in terms of precision and recall.
Also, the authors reported that only a limited number of refactoring instances could not be detected without a specific similarity. In other words, the detected refactoring instances were mostly common between the three types of similarities.

\subsection{Metrics-Based Approach}
The approach based on metrics detects refactoring instances by difference in metric values between the two versions.
This approach detects instances rapidly because it adopts a lightweight analysis compared to the rule-based approach described in Section~\ref{sec:rule}.
However, this rapid approach has the low accuracy in refactoring detection because it does not analyze source code syntactically.
The technique described by Demeyer \etal\ selects metrics such as method and class size and inheritance, from metrics of Chidamber \& Kemerer~\cite{chidamber-tse1994} and Lorenz \& Kidd~\cite{lorenz-book1994} and then uses combinations of these metrics as heuristics to detect refactoring instances such as splitting of methods, and merging and splitting of child classes or parent classes~\cite{demeyer-oopsla2000}.

Moreover, Mahouachi \etal\ also proposed a search-based technique that detects refactoring based on the differences in structural metrics between two versions~\cite{mahouachi-ssbse2013}.
Their technique is built around a search-based process that minimizes the difference in metrics using a genetic algorithm. Their approach also is described in Section \ref{s:search-based}.

\subsection{Dynamic Analysis-Based Approach}
Research have also been done on detection of refactoring with attention to the assumption of maintaining program behavior while refactoring.
The dynamic analysis-based approach verifies that program behavior is maintained by executing test cases with related modifications and examining the consistency of the results.
This approach only enables detecting \textit{pure refactoring} without any \textit{impure refactoring}.
On the other hand, a disadvantage of the dynamic analysis-based approach is that it is impossible to identify the kind of refactoring performed from the information in the dynamic analysis.

Soares \etal\ proposed a refactoring detection technique using a tool called \textsf{SafeRefactor}~\cite{soares-ieee-softw-2010}, to detect changes in program behavior while refactoring~\cite{soares-sbes-2011}.
\textsf{SafeRefactor} automatically generates unit tests for non-changed methods among versions.
Then, it executes the generated tests and identifies any changes in behavior shown by failing the tests.
Their technique inputs an original version of the source code and a modified version to \textsf{SafeRefactor} and identifies refactoring if \textsf{SafeRefactor} verifies program behavior is maintained.

\subsection{Graph Matching-Based Approach}\label{s:matching-based}

Some researchers have proposed techniques to detect refactoring instances by regarding a program or a program change as a graph structure and by checking whether patterns of refactoring operations are included in the graph as a subgraph of it.
Since most software design models, such as UML class diagrams, can be regarded as graphs, it is straightforward to use graph matching to detect model refactorings.
Handling refactorings to be detected as patterns leads to simplicity in the definition of the detection mechanism.
In addition, handling a code change as a graph enables us to detect refactorings as a subgraph even when mixed with other changes.
However, there are several disadvantages to this approach such as the difficulty of defining patterns for complicated models and for some refactoring types.

Kehrer \etal\ proposed a technique to increase the abstraction level of a difference in models by extracting high-level changes such as refactoring operations from an operation sequence of EMF models, i.e., a model difference~\cite{kehrer-ase2011}.
Changes among versions in a model can be expressed as primitive operations in a graph such as additions or deletions of nodes or edges.
High-level changes included in such a graph are then detected as a subgraph, and grouped to identify a higher abstraction level representation of the change.
Kehrer \etal\ called such grouping manipulation semantic lifting and automated this process.

Soetens \etal\ proposed a detection technique of \textit{floss refactoring} based on matching of an operation history~\cite{soetens-icsm2013}.
Since such \textit{floss refactoring} is performed together with other changes, information obtained from the versions may be mixed, containing multiple modifications, which makes it difficult to detect refactoring instances in it.
In their technique, a graph representing code changes is constructed based on the edit operation history of the source code as recorded by a tool named \textsf{ChEOPSJ}~\cite{soetens-csmr12}.
This technique then confirms whether this graph contains subgraph patterns representing refactoring operations using a graph transformation tool.
If it contains patterns, then the technique detects the corresponding refactorings that have been performed.
The authors claimed that refactoring patterns such as \textit{Rename Method} and \textit{Move Method} can be detected more accurately using operation histories than by other existing techniques.

\subsection{Search-Based Approach}\label{s:search-based}
It is important to properly detect compound refactoring and floss refactoring, refactoring operations mixed together with other changes.
When \textit{impure refactoring} is performed, the changes made between versions before and after the refactoring session are affected by not only the single refactoring, but also by other refactoring and/or non-refactoring operations.
In such cases, refactoring instances cannot be correctly detected by only looking for the pre- and post-conditions of a single refactoring instance because the difference between versions before and after the changes will not correspond to the conditions.

In Search-Based Software Engineering~(SBSE)~\cite{SBSE}, a software engineering problem is regarded as a sort of optimization problem, and the results are obtained using search techniques.
There are several applications of SBSE in refactoring detection techniques.
In search-based refactoring detection, a program and a refactoring application are respectively regarded as a state and an operator of the state transition, and an optimal sequence of operators representing the changes in the program between versions is discovered.
The search progresses by repeatedly invoking refactoring of the program as an operator application, obtaining a new program.
On the one hand, the search-based approach has the advantage that it does not require detection rules for \textit{impure refactoring} directly because it can indirectly handle intermediate program states where only some of the mixed changes were applied.
On the other hand, the disadvantages of the search-based approach include that some search techniques require a large computational time.

As an example of this approach, P\'{e}rez and Crespo proposed a search-based technique to identify refactoring operations from the structural differences in a program, such as changes in a UML class diagram~\cite{perez-evol2007}.
In this technique, a depth-first search is applied to find a sequence of refactorings, and invoking the detected refactoring candidates on the program.

Hayashi \etal\ proposed a technique to detect refactoring operations using the \textit{A} search~\cite{hayashi-apsec2008,hayashi-ieicet2010}.
In this technique, refactoring detection is formulated as a path search problem, regarding the size of the structural differences in the program as a heuristic distance, the weighted count of the applied refactorings as the path distance, and the sum of these as the evaluation function.
The solution path is then discovered using the \textit{A} search.

This approach also includes applications of the genetic algorithm.
In this technique, a refactoring sequence is represented as a chromosome consisting of multiple genes.
The algorithm then finds an optimal chromosome, i.e., a chromosome which has the maximum value of the fitness function, as the most appropriate sequence of refactorings explaining the changes between versions, by iteratively applying the updating operators such as selection, crossover, and mutation.
Fadhel \etal\ proposed a technique to detect model refactorings using the genetic algorithm~\cite{fadhel-icsm2012}.
Mahouachi \etal\ proposed a technique applying a similar approach to source code to obtain a sequence of code refactorings~\cite{mahouachi-ssbse2013}.
In this technique, a fitness function is defined as minimizing the differences in product metrics between versions.

Thangthumachit \etal\ proposed a technique to detect refactoring operations based on the similarity of child and referencing elements in an abstract syntax tree~\cite{zui-apsec2011}.
In this technique, refactoring detection is performed at each level, such as package, file, class, and method, and refactorings detected at a coarser-grained level are applied to the target program before trying to detect refactorings at a finer-grained level.
Refactorings that failed to apply were excluded from the detection result.
By repeating the detection and the application in this manner, an accurate detection result is achieved even for a difference in which multiple refactoring operations are mixed.
In addition, a tool has also been proposed to visualize the sequence of refactorings obtained in this way~\cite{hayashi-wcre2013}.

\section{Future Directions}
\label{sec:discussion}
\subsection{Combination of Multiple Techniques}

In future work, first, techniques for refactoring detection should be combined and  evaluated.
Soares \etal\ suggested a quantitative comparison of combined techniques for refactoring detection as future work in their paper on a quantitative comparison~\cite{Soares2013}.

As can be seen in Table~\ref{tab:category}, very little research has been done on combined approaches so far.
For example, as future work, a search-based approach could be combined not only with rule-based approach that considers program structures, but also with a clone detection-based approach that identifies moving of code fragments between versions.
Then these combined approaches should be compared with other existing techniques.
Similarly, very little research has been done on dynamic analysis-based detection.
Since a static analysis-based detection approach, such as code clone analysis-based approach, cannot identify a change in external behavior, it should be combined with a dynamic analysis-based approach.

\subsection{Quantitative Evaluation}
Second, very little research has been done on the quantitative comparison of refactoring detection techniques.
One reason is that it is difficult to define refactoring instances that should be detected.
To enable such comparison, the research community should provide datasets to be used for quantitative comparison, and then set up a mechanism to share the datasets.
Soares \etal\ performed a quantitative comparison of \textsf{Ref-Finder}~\cite{kim-fse2010,prete-icsm2010}, a dynamic analysis-based technique, and a commit-log-based technique, and then published the results on their website~\cite{Soares2013}. 
To improve the techniques, it is necessary to compare the various refactoring detection techniques and publish the comparison results.
In a dataset used for this comparison, the definition of the refactoring instances should be clear, e.g., whether a dataset includes \textit{impure}/\textit{floss refactoring} instances or not.
In addition to the quantitative evaluation of detection performance, quantitative evaluation of scalability with respect to the number of revisions is also needed.
For a large-scale empirical study of refactoring, a scalable approach is needed that completes the detection process within a practical duration by analyzing only differences and associated code among revisions.
Since most of the existing techniques are aimed at detecting refactoring instances between a revision pair by analyzing all of the source code of each revision, analyzing a longitudinal sequence of revisions requires a large computational cost.
As future work, then, once the quantitative evaluation of scalability with respect to the number of revisions is performed, a scalable tool needs to be developed.

\section{Conclusion}\label{sec:summary}

In this paper, we surveyed refactoring detection techniques mainly focuses on analysis of change histories, which has been well studied recently.
First, we explained the definition of the refactoring detection terms in this paper, and classified refactoring detection techniques into four categories: mining commit logs, observing developers, analyzing tool usage logs, and analyzing change histories.
Next, we classified the techniques based on change history analysis into six subcategories, and we introduced the techniques belonging to each subcategory.
Finally, we discussed two directions for future research, combinations of multiple techniques, and quantitative evaluation of the techniques.

We hope that this paper will help encourage further improvements in refactoring detection techniques.

\section*{Acknowledgments}
This work was partially supported by MEXT/JSPS KAKENHI Grant Numbers 26730036, and 23700030.

\end{document}